# PROPERTIES OF THE FINANCIAL BREAK-EVEN POINT IN A SIMPLE INVESTMENT PROJECT AS A FUNCTION OF THE DISCOUNT RATE

Domingo Alberto Tarzia

Depto. Matemática – CONICET
FCE, Univ. Austral
Paraguay 1950, S2000FZF Rosario, ARGENTINA.
E-mail: DTarzia@austral.edu.ar

*Abstract.* We consider a simple investment project with the following parameters: $I > 0$: Initial outlay which is amortizable in $n$ years; $n$: Number of years the investment allows production with constant output per year; $A > 0$: Annual amortization ($A = I/n$); $Q > 0$: Quantity of products sold per year; $C_v > 0$: Variable cost per unit; $p > 0$: Price of the product with $p > C_v$; $C_f > 0$: Annual fixed costs; $t_e$: Tax of earnings; $r$: Annual discount rate. We also assume inflation is negligible.

We derive a closed expression of the financial break-even point $Q_f$ (i.e. the value of $Q$ for which the net present value ($NPV$) of the investment project is zero) as a function of the parameters $I$, $n$, $C_v$, $C_f$, $t_e$, $r$, $p$. We study the behavior of $Q_f$ as a function of the discount rate $r$ and we prove that: (i) For $r$ negligible $Q_f$ equals the accounting break-even point $Q_c$ (i.e. the earnings before taxes (EBT) is null); (ii) When $r$ is large the graph of the function $Q_f = Q_f(r)$ has an asymptotic straight line with positive slope. Moreover, $Q_f(r)$ is an strictly increasing and convex function of the variable $r$; (iii) From a sensitivity analysis we conclude that, while the influence of $p$ and $C_v$ on $Q_f$ is strong, the influence of $C_f$ on $Q_f$ is weak; (iv) Moreover, if we assume that the output grows at the annual rate $g$ the previous results still hold, and, of course, the graph of the function $Q_f = Q_f(r,g)$ *vs r* has, for all $g > 0$, the same asymptotic straight line when $r \to \infty$ as in the particular case with *g=0*.

## 1. Introduction

A rigorous evaluation of an investment project is crucial for the evaluation of its convenience, see the books (Alhabeeh, 2012; Bierman and Smidt, 1993; Bodmer, 2015; Brealey and Myers, 1993; De Pablo, Ferruz and Santamaria, 1990; Lopez Dumrauf, 2003; Suarez Suarez, 1991). Among the different methods to evaluate investment project we choose the net present value ($NPV$) (Baker and Fox, 2003; Beaves, 1988; Bric and Weaver, 1997; Chung and Lin, 1998; Grinyer and Walker, 1990; Hadjdasinski (1993, 1995, 1996, 1997); Hartman and Schafrick, 2004; Hazen, 2003; Kim and Chung, 1990; Lan, Chung, Chu and Kuo, 2003; Lohmann, 1994; Lohmann and Baksh, 1993; Moon and Yun, 1993; Pasin and Leblanc, 1996; Pierru and Feuillet Midrier, 2002; Prakash, Dandapani and Karels, 1988; Reichelstein, 2000; Roumi and Schnabel, 1990; Shull, 1992; Stanford, 1989; and Zhang (2005)). Assuming the absence of inflation we consider a simple investment project with the following parameters:

$I > 0$: Initial outlay which is amortizable in $n$ years;

$n > 1$: Numbers of years of the explicit forecasted period of the investment project which make the same activities per year with only one product;

$Q > 0$: Quantity of products sold per year;

$C_v > 0$: Variable cost per unit;

$p > 0$: Price per unit with $p > C_v$;

$C_f > 0$: Annual fixed costs;

$t > 0$: Time;

$t_e > 0$: Tax of earnings (legal tax rate);

$r > 0$: Annual discount rate;

$g > 0$: Annual growth rate.

Moreover, we also consider the annual amortization ($A = I/n > 0$) which depends of two parameters $I$ and $n$.

In Fernandez Blanco (1991) a first study for the $NPV$ of an investment project was done, and we complete and improve it. There exist several papers on $NPV$ but, from our point of view, we have not found in the literature a study of the mathematical-financial properties of the financial break-even point and this is the main objective of the present paper. We derive first an explicit expression of the $NPV$ as a function of the independent variable $Q$ in order to obtain a closed formula of the financial break-even point $Q_f$ (the value of $Q$ for which the $NPV$ of the investment project is zero) as a function of the parameters ($I$, $n$, $C_v$, $C_p$, $t_e$, $r$, $p$).

Recent applications of the net present value and the break-even analysis can be found: break-even point between short-term and long-term capital gain (loss) strategies, Yang and Meziani, 2012; break-even procedure for a multi-period project, Kim and Kim, 1996; to mazimize the net present value of projects, Schwindt and Zimmermann, 2001; Vanhoucke, Demeulemeester and Herroelen, 2001a,b; the internal rate of return as a financial indicator, Hajdasinski, 2004; private competitiveness, production costs and break-even analysis of representative production units, Martinez Medina et al., 2015; the net present value as an optimal criterium of investing, Machain, 2002; break-even points for storage systems as a substitute to conventional grid reinforcements, Nykamp et al. 2014; to measure and analyse the operating risk, financial risk, financial break-even point and total risk of a selected public sector, Sarkar and Sarkar, 2013; risk diagnosis in the context of economic crisis, Suciu, 2010; marginal break-even between maintenance strategies alternatives, Gokiene, 2010; relationship among discount cash flow, free cash flow, economic value added and net present value, Hartman, 2010; Shieres and Wachowicz, 2001; running a profitable company, Paek, 2000;

The break-even analysis is a useful tool to study the relationship among fix costs, variable costs and returns. The break-even analysis computes the volume of production at a given price necessary to cover all costs. We study the behavior of $Q_f = Q_f(r)$ with respect of the discount rate $r$ and we prove the following results:

(i) When $r$ is negligible ($r$ goes to zero) then $Q_f$ tends to the accounting break-even point $Q_c$ (the value of $Q$ for which the earnings before taxes (EBT) of the investment project is equals to zero);

(ii) When $r$ is large ($r$ goes to infinity) the graph of the function $Q_f = Q_f(r)$ has an asymptotic straight line. Its positive slope and the y-intercept point at $r=0$ are determined explicitly. Moreover, $Q_f(r)$ is an strictly increasing and convex function of the variable $r$;

(iii) By a sensitivity analysis we obtain that $p$ and $C_v$ have an important influence on $Q_f$, but $C_f$ has a negligible influence on $Q_f$;

(iv) Moreover, if we assume that the output grows at the annual rate $g$ the previous results still hold and, of course, the graph of the function $Q_f = Q_f(r,g)$ vs $r$ has, for all $g > 0$, the same asymptotic straight line when $r \to \infty$ as in the particular case with $g=0$.

## 2. Investment Project with the Independent Variable Q

We assume that in each year ($t = 1, 2, ..., n$) the company do the same activities, i.e. the parameters $Q$, $p$, $C_f$, $C_v$, $r$, $t_e$ are constants during the $n$ years of the investment project. For each year $t$ ($t = 1, 2, ..., n$) the net cash flow is given by:

$$\begin{aligned} F &= (1-t_e)\left[(p-C_v)Q - C_f - A\right] + A \\ &= (1-t_e)\left[(p-C_v)Q - C_f\right] - (1-t_e)A + A \\ &= (1-t_e)\left[(p-C_v)Q - C_f\right] + t_e A, \end{aligned}$$

and then, by using the calculation of the sum of the first n terms of a geometric sequence, the $NPV$ of the investment project is given by (Brealey and Myers, 1993; Sapag Chain, 2001; Tang and Tang, 2003; Villalobos, 2001):

$$NPV(Q) = -I + \sum_{t=1}^{n} \frac{F}{(1+r)^t} = -I + F \frac{1}{r}\left[1 - \frac{1}{(1+r)^n}\right]$$

$$= -I + \frac{1}{r}\left[1 - \frac{1}{(1+r)^n}\right]\left[(1-t_e)(p-C_v)Q - (1-t_e)C_f + t_e A\right] = h + mQ \quad (1)$$

which represents a straight line of the variable $Q$ with

$$h = h(r) = -I + f(r)\left[t_e A - (1-t_e)C_f\right] \text{ (y-intercept of the straight line)}, \qquad (2)$$

$$m = m(r) = f(r)(p - C_v)(1 - t_e) > 0 \text{ (slope of the straight line)}, \qquad (3)$$

where the real function $f = f(r)$ is defined by the following expression:

$$f(r) = \frac{1}{r}\left[1 - \frac{1}{(1+r)^n}\right], \quad r > 0. \qquad (4)$$

Taking into account that the financial break-even $Q_f$ is defined as the value of $Q$ which satisfies $NPV(Q) = 0$, we obtain that

$$NPV(Q_f) = 0 \iff h(r) + m(r)Q_f = 0, \qquad (5)$$

that is,

$$Q_f = Q_f(r) = \left\{(1-t_e)C_f - t_e A + \frac{I}{f(r)}\right\}\frac{1}{(p-C_v)(1-t_e)} = -\frac{h(r)}{m(r)}, \qquad (6)$$

and therefore the financial break-even point $Q_f$, as a function of the discount rate $r$, is given by the following expression:

$$Q_f(r) = a + b\frac{1}{f(r)}, \qquad (7)$$

where the real coefficients $a$ and $b$ are given by

$$a = a = \frac{C_f - t_e(C_f + A)}{(p - C_v)(1 - t_e)} \qquad (8)$$

$$b = \frac{I}{(p - C_v)(1 - t_e)} > 0. \qquad (9)$$

Taking into account the formula (6) of the financial break-even point $Q_f$ the formula (7) of the net present value $NPV(Q) = NPV(Q, r)$ has an equivalent expression given by

$$NPV(Q, r) = -I + f(r)\left[t_e A - (1-t_e)C_f\right] + f(r)(p - C_v)(1 - t_e)Q, \qquad (10)$$

or equivalently by

$$NPV(Q, r) = m(r)\left[Q - Q_f(r)\right]. \qquad (11)$$

Thus we have obtained an expression of the $NPV$ as a function of the variable $Q$, the discount rate $r$, and the financial break-even point $Q_f(r)$. The previous result can be summarized as follows.

*Theorem 1*

The investment project has the following properties:

(i) The $NPV$, as a function of the units sold per year $Q$, is given by (1) where the y-intercept $h$ and the slope $m$ are expressed by (2) and (3) respectively where $f = f(r)$ is the real function defined by (4).

(ii) The financial break-even point $Q_f$, as a function of the discount rate $r$, is given by

$$Q_f(r) = a + b\, F(r), \tag{12}$$

where the real function $F = F(r)$ is defined by

$$F(r) = \frac{1}{f(r)} = \frac{r}{1 - \frac{1}{(1+r)^n}}, \quad r > 0, \tag{13}$$

and the coefficients $a$ and $b$ are given by the expressions (8) and (9) respectively.

(iii) The $NPV(Q,r)$ can also be calculated as a function of the $Q_f$ by the expression (11).

(iv) The sign of the $NPV(Q,r)$, as a function of the financial break-even point $Q_f(r)$, is given by

$$NPV(Q,r) \begin{cases} > 0 \Leftrightarrow Q > Q_f(r) \\ = 0 \Leftrightarrow Q = Q_f(r) \\ < 0 \Leftrightarrow 0 \leq Q < Q_f(r). \end{cases} \tag{14}$$

In order to analyse the matematical behavior of the function $NPV = NPV(Q,r)$ and the financial break-even point $Q_f = Q_f(r)$ we need to study the behavior of the real functions $f = f(r)$ and $F = F(r)$ defined by (4) and (13) respectively which have the following properties.

*Theorem 2*

(i) The function $f = f(r)$ is a strictly decreasing and convex function of the discount rate $r$ with the following properties:

$$f(0^+) = \lim_{r \to 0^+} f(r) = n > 0, \quad f(+\infty) = \lim_{r \to +\infty} f(r) = 0, \tag{15}$$

$$\frac{df(r)}{dr} = f'(r) = -\frac{G(r)}{r^2(1+r)^{n+1}} < 0, \quad \forall r > 0, \tag{16}$$

$$f'(0^+) = \lim_{r \to 0^+} f'(r) = -\frac{n(n+1)}{2}, \quad f'(+\infty) = \lim_{r \to +\infty} f'(r) = 0, \tag{17}$$

$$f''(r) = \frac{H(r)}{r^3(1+r)^{n+2}} > 0, \quad \forall r > 0, \tag{18}$$

$$f''(0^+) = \lim_{r \to 0^+} f''(r) = \frac{n(n+1)(n+2)}{3}, \quad f''(+\infty) = \lim_{r \to +\infty} f''(r) = 0, \tag{19}$$

where the real functions $G = G(r)$ and $H = H(r)$ are defined by

$$G(r) = (1+r)^{n+1} - 1 - (n+1)r, \ r > 0, \tag{20}$$

$$H(r) = 2(1+r)^{n+2} - 2 - 2(n+2)r - (n+1)(n+2)r^2, \ r > 0, \tag{21}$$

which have the following properties:

$$G(0^+) = \lim_{r \to 0^+} G(r) = 0, \ G(+\infty) = \lim_{r \to +\infty} G(r) = +\infty, \ G(r) > 0, \ \forall r > 0, \tag{22}$$

$$H(0^+) = \lim_{r \to 0^+} H(r) = 0, \ H(+\infty) = \lim_{r \to +\infty} H(r) = +\infty, \ H(r) > 0, \ \forall r > 0. \tag{23}$$

(ii) The real function $F = F(r)$, defined in (13), is a strictly increasing function and has at $r = +\infty$ an asymptotic straight line given by the equation $y = r$ (straight line with slope 1 and y-intercept 0) and the following properties:

$$F(0^+) = \lim_{r \to 0^+} F(r) = \frac{1}{n}, \ F(+\infty) = \lim_{r \to +\infty} F(r) = +\infty, \tag{24}$$

$$\frac{1}{2} < F'(0^+) = \lim_{r \to 0^+} F'(r) = \frac{1}{2}\left(1 + \frac{1}{n}\right) < 1, \ \forall n > 1, \tag{25}$$

$$F''(0^+) = \lim_{r \to 0^+} F''(r) = \frac{n^2 - 1}{6n}, \tag{26}$$

$$0 < F(r) - r < \frac{1}{n}, \ \forall r > 0, \ \forall n > 1. \tag{27}$$

*Proof.*

All properties of the real functions $f, F, G$ and $H$ can be proved by using elementary mathematical analysis (derivatives, l'Hopital rule, increasing and convexity of functions, asymptotic straight lines, etc.). ∎

Taking into account that the earnings before taxes (*EBT*) is calculated by

$$EBT = pQ - C_v Q - C_f - A = (p - C_v)Q - C_f - A,$$

and defining the accounting break-even point $Q_c$ as the value $Q$ for which the *EBT* is zero then $Q_c$ is given by the following expression

$$BAT(Q_c) = 0 \iff (p - C_v)Q_c - C_f - A = 0 \iff$$

$$Q_c = \frac{C_f + A}{p - C_v}. \tag{28}$$

### Theorem 3

The financial break-even point $Q_f = Q_f(r)$, given by (12), is a strictly increasing function of the discount rate $r$ and has the following properties:

$$Q_f\left(0^+\right) = \lim_{r \to 0^+} Q_f(r) = a + \frac{b}{n} = Q_c, \quad Q_f(+\infty) = \lim_{r \to +\infty} Q_f(r) = +\infty, \tag{29}$$

$$\frac{dQ_f(r)}{dr} > 0, \quad \forall r > 0, \tag{30}$$

$$0 < \frac{b}{2} < \frac{dQ_f(0^+)}{dr} = \lim_{r \to 0^+} \frac{dQ_f}{dr}(r) = \frac{b}{2}\left(1 + \frac{1}{n}\right) < b. \tag{31}$$

Moreover, the function $y = Q_f(r)$ has at $r = +\infty$ a straight line given by the equation

$$y = a + b\, r, \tag{32}$$

which has a slope $b > 0$ and y-intercept $a$, defined in (9) and (8) respectively.

*Proof*

Taking into account the properties of the functions $f = f(r)$ and $F = F(r)$, obtained in Theorem 2, we have the following results:

$$Q_f(0^+) = \lim_{r \to 0^+} Q_f(r) = a + b \lim_{r \to 0^+} F(r) = a + bF(0^+) = a + \frac{b}{n} = \frac{C_f + A}{p - C_v} = Q_c \tag{33}$$

which is the accounting break-even point $Q_c$ defined by (28). On the other words, we have the following properties:

$$Q_f(+\infty) = \lim_{r \to +\infty} Q_f(r) = a + b \lim_{r \to +\infty} F(r) = a + bF(+\infty) = +\infty, \tag{34}$$

$$\frac{dQ_f}{dr}(r) = bF'(r) > 0, \quad \forall r > 0, \tag{35}$$

$$\frac{b}{2} < \frac{dQ_f}{dr}(0^+) = bF'(0) = \frac{b}{2}\left(1 + \frac{1}{n}\right) < b, \quad \forall n > 1. \tag{36}$$

Moreover, the function $y = F(r)$ for $r = +\infty$ becomes asymptotic to the straight line whose equation is given by the equation $y = r$, and then the function $y = Q_f(r)$ for $r = +\infty$ is asymptotic to the straight line $y = a + br$ because

$$\text{i) } \lim_{r \to +\infty} \frac{Q_f(r)}{r} = \lim_{r \to +\infty} \frac{a + bF(r)}{r} = b \lim_{r \to +\infty} \frac{F(r)}{r} = b,$$

$$\text{ii) } \lim_{r \to +\infty} \left[Q_f(r) - br\right] = \lim_{r \to +\infty} \left[a + bF(r) - br\right] = a + b \lim_{r \to +\infty} \left(F(r) - r\right) = a. \quad \blacksquare \tag{37}$$

*Remark 1*

The limit (33) has an interesting accounting-financial property: the limit of the financial break-even point $Q_f = Q_f(r)$ when the discount rate goes to zero (i.e., discount rate is negligible) is equal to the accounting break-even point $Q_c$. ∎

*Remark 2*

The financial break-even point of the investnment project, as a function of the discount rate $r$, is given by a strictly increasing and convex function $y = Q_f(r)$ which at $r = 0$ has the value $Q_f(0^+) = a + b/n = Q_c$ (accounting break-even point) and for $r = +\infty$ tends asymptotically to the straight line of equation $y = a + br$ where $a$ and $b$ are defined in (8) and (9) respectively. On the other hand, the curve $y = Q_f(r)$ has an initial slope $Q_f'(0^+)$ at $r = 0$ which has a value between $b/2$ and $b$, less than the slope $b$ of the asymptote for $r = +\infty$. ∎

Taking into account the result (29) an interesting question is to determine the rate of convergence of the financial break-even point $Q_f = Q_f(r)$ to the accounting break-even point $Q_c$ when the discount rate $r$ goes to zero.

*Lemma 4*

We have

$$0 < Q_f(r) - Q_c = \frac{I}{(p - C_v)(1 - t_e)}\left(F(r) - \frac{1}{n}\right), \quad \forall r > 0, \tag{38}$$

and the rate of convergence is of order one and it is given by

$$0 < Q_f(r) - Q_c \approx \frac{I}{2(p - C_v)(1 - t_e)}\left(1 + \frac{1}{n}\right)r, \quad \text{as } r \to 0. \tag{39}$$

*Proof*

Taking into account formulas (7) or (12) for $Q_f(r)$, and formula (28) for $Q_c$, we obtain:

$$Q_f(r) - Q_c = \frac{C_f - t_e(C_f + A) + I F(r)}{(p - C_v)(1 - t_e)} - \frac{I F(r) - A}{p - C_v} = \frac{I F(r) - A}{(p - C_v)(1 - t_e)},$$

that is (38). Now, taking into account the properties (24) and (25) for the function $F(r)$ we get (39). ∎

Now we will study the $NPV$ as a function of the two independent variables: the discount rate $r$ and the number $Q$ of units sold per year.

*Theorem 5*

(i) The $NPV$, as a function of the two independent variables $Q$ and $r$, is given by the following expression

$$NPV(Q, r) = (p - C_v)(1 - t_e) f(r)(Q - a) - I \tag{40}$$

where the function $f = f(r)$ was defined in (4).

(ii) The $NPV(Q,r)$ is a strictly increasing function of the variable $Q$ and a strictly decreasing function of the variable $r$ which has for the extreme values $0$ and $+\infty$ of the variables $Q$ and $r$ respectively the following expressions:

$$NPV(Q,+\infty) = \lim_{r \to +\infty} NPV(Q,r) = -I < 0, \quad \forall Q > 0, \tag{41}$$

$$NPV(Q,0^+) = \lim_{r \to 0^+} NPV(Q,r) = n(p-C_v)(1-t_e)(Q-Q_c), \quad \forall Q > 0, \tag{42}$$

$$NPV(0^+,r) = \lim_{Q \to 0^+} NPV(Q,r) = h(r), \quad \forall r > 0, \tag{43}$$

$$NPV(+\infty,r) = \lim_{Q \to +\infty} NPV(Q,r) = +\infty, \quad \forall r > 0, \tag{44}$$

where $Q_c$ is the accounting break-even point defined in (28), and $h = h(r)$ is the real function defined in (2).

(iii) The real function $h = h(r)$ has the following properties:

$$h(0^+) = -(1-t_e)n(A+C_f) < 0, \; h(+\infty) = -I < 0, \tag{45}$$

and it is a strictly increasing (decreasing) function when $At_e < C_f(1-t_e)$ $\left(At_e > C_f(1-t_e)\right)$.

In the particular case in with $At_e = C_f(1-t_e)$, then $h = h(r)$ is a constant function given by $h(r) = -I < 0, \forall r > 0$.

*Proof.*

The properties follow from the previous results and elemenarty mathematical computations.

In particular, the partial derivatives of $NPV(Q,r)$ with respect to the variables $Q$ and $r$ are given by the following expressions:

$$\frac{\partial NPV}{\partial Q}(Q,r) = (p-C_v)(1-t_{ig})f(r) > 0, \quad \forall Q, r > 0, \tag{46}$$

$$\frac{\partial NPV}{\partial r}(Q,r) = (p-C_v)(1-t_{ig})Q f'(r) < 0, \quad \forall Q, r > 0, \tag{47}$$

and then the properties (i) and (ii) hold.

Therefore, the derivative of $h = h(r)$ is given by:

$$h'(r) = \left[C_f(1-t_e) - At_e\right]\frac{G(r)}{r^2(1+r)^n}, \quad \forall r > 0, \tag{48}$$

where $G = G(r)$ was defined in (20). The sign of $h'(r)$ depends of the sign of $\left[C_f(1-t_e) - At_e\right]$, which will be positive (i.e. $h$ is a strictly increasing function of the variable $r$) when $At_e < C_f(1-t_e)$ and so on. In the particular case $At_e = C_f(1-t_e)$ we get that $h'(r) = 0, \forall r > 0$, and then $h(r)$ is a constant given by $h(r) = -I, \forall r > 0$. ∎

*Remark 3*

In the Theorem 5 we showed that the increasing or decreasing behavior of the real function $h = h(r)$ depends on the sign of the expression

$$C_f(1-t_e) - At_e, \qquad (49)$$

which has a financial-accounting interpretation. ∎

*Remark 4:*

Owing to the properties (46) and (47) of the $NPV(Q,r)$, as a real function of the two independent variables $Q$ and $r$, and the expression (11) of the $NPV$ with parameters $I$, $n$, $p$, $C_v$, $A = I/n$, $C_f$, $t_e$, the zero curve of the $NPV = NPV(Q,r)$ is given by:

$$NPV(Q,r) = 0 \iff (p - C_v)(1 - t_{ig}) f(r)(Q - a) = I \iff \qquad (50)$$

$$Q = a + \frac{I}{(p - C_v)(1 - t_{ig}) f(r)} = a + \frac{b}{f(r)} = Q_f(r)$$

and therefore, the zero curve of $NPV(Q,r)$ is given, in the plane $Q$, $r$, by the equation:

$$Q = Q_f(r), \quad \forall r > 0, \qquad (51)$$

or equivalently by

$$r = Q_f^{-1}(Q), \quad Q > Q_c, \qquad (52)$$

where $Q_f(r)$ is the financial break-even point for the quantity $Q$ as a function of the discount rate $r$, defined by (12), and $Q_f^{-1}$ is the inverse function of $Q_f$ which is defined $\forall Q > Q_c$, where $Q_c$ is the accounting break-even point given by (28). ∎

## 3. Numerical Results for an Investment Project

We will consider the investment project with the following parameters (Brealey and Myers (1993)):

- initial outlay: $I = 150000 \; (\$)$;
- explicit forecasted period of the project: $n = 10$;
- annual amortization: $A = 15000 \; (\$)$;
- price per unit: $p = 3.70 \; (\$/unit)$;
- variable cost per unit: $C_v = 3.00 \; (\$/unit)$;
- annual fixed costs: $C_f = 30000 \; (\$)$;
- tax of earnings: $t_e = 0.35$ (35 % annual);
- discount rate: $r = 0.10$ (10 % annual);

Owing to the previous theoretical results we have:

$$a = \frac{C_f - t_e(C_f + A)}{(p - C_v)(1 - t_e)} = \frac{30000 - 0.35(30000 + 15000)}{(3.70 - 3)0.65} = 31318.68 \text{ (Units)}$$

$$b = \frac{I}{(p - C_v)(1 - t_e)} = \frac{150000}{(3.70 - 3)0.65} = 329670,33 \text{ (Units)}$$

$$Q_c = \frac{C_f + A}{p - C_v} = a + \frac{b}{n} = \frac{30000 + 15000}{3.70 - 3} = \frac{45000}{0.70} = 64285.71 \text{ (Units)}$$

$$t_e A - (1 - t_e)C_f = -C_f + t_e(C_f + A) = -14250 \text{ (\$)}$$

$$f = f(0.10) = \frac{1}{0.10}\left[1 - \frac{1}{(1.10)^{10}}\right] = 6.14$$

$$h = h(0.10) = -I + f(0.10)\left[t_e A - (1 - t_e)C_f\right] = -237560.08 \text{ (\$)}$$

$$m = m(r) = f(r)(p - C_v)(1 - t_{ig}) = 2.80 \text{ (\$/Units)}$$

$$Q_f = Q_f(0.10) = a + \frac{b}{f(0.10)} = -\frac{h(0.10)}{m(0.10)} = 84971.01 \text{ (Units)}$$

$$NPV(Q) = h + mQ = 2.80(Q - 84971.01) \text{ (\$)}.$$

Taking into account the same parameters given before we can obtain the values of the financial break-even point $Q_f(r)$ as a function of the discount rate $r$ (see Table 1), and then we can get the graph of the real function $Q_f(r)$ vs $r$ (see Figure 1). Moreover, we can also plot the financial break-even point $y = Q_f(r)$ and the asymptotic straight line $y = a + b\,r$ as a function of the discount rate $r$ (see Figure 2).

Now we will perform a sensitivity analysis of the investment project in a neighboord of a reference point $r = 0.10$. We get the sensitivity analysis of the financial break-even point $Q_f(r)$ with respect to the price $p$, the annual fixed costs $C_f$ and the variable cost per unit $C_v$ as a function of the discount rate $r$. For that, we compute $Q_f(r)$ for the discount rate $r = 0.11\,(+10\%)$, $r = 0.12\,(+20\%)$, and $r = 0.09\,(-10\%)$, and we compare the results with those for the reference discount rate $r = 0.10$ (see Figure 3).

The sensitivity analysis of the financial break-even point $Q_f(r)$ suggests that price per unit $p$ lies on a more sensible point in comparison to the costs of any kind. If we look into the costs, it is seen that annual fixed costs $C_f$ is less sensible than variable costs per unit $C_v$ against the discount rate $r$

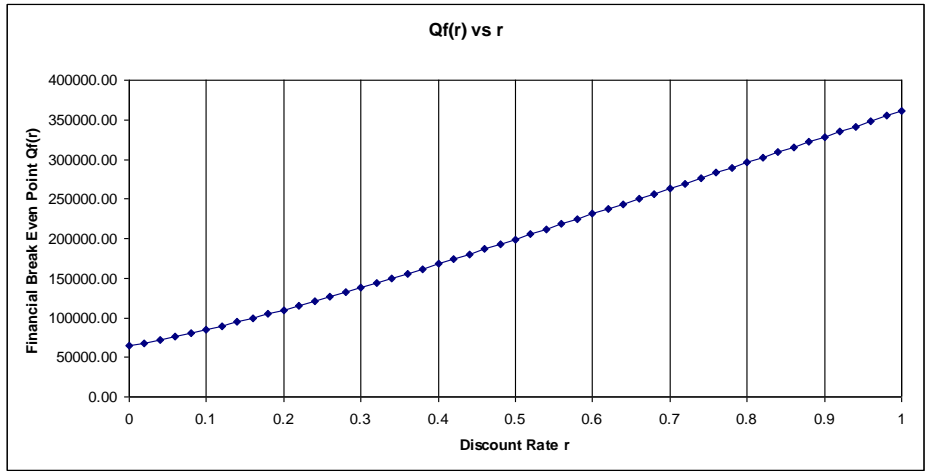

*Figure 1 Graph of* $Q_f(r)$ *vs.* $r$

| Table 1 Values of the financial break-even point $Q_f(r)$ vs the discount rate $r$ | Discount rate r | Financial break-even point $Q_f(r)$ |
|---|---|---|
| | 0.03 | 69966.10 |
| | 0.04 | 71964.05 |
| | 0.05 | 74012.50 |
| | 0.06 | 76110.32 |
| | 0.07 | 78256.32 |
| | 0.08 | 80449.28 |
| | 0.09 | 82687.94 |
| | 0.10 | 84971.01 |
| | 0.11 | 87297.17 |
| | 0.12 | 89665.11 |
| | 0.13 | 92073.48 |
| | 0.14 | 94520.95 |
| | 0.15 | 97006.17 |
| | 0.16 | 99527.83 |
| | 0.17 | 10208.59 |
| | 0.18 | 104675.16 |
| | 0.19 | 107298.23 |
| | 0.20 | 109952.56 |
| | 0.25 | 123650.30 |
| | 0.30 | 137954.98 |
| | 0.35 | 152742.30 |
| | 0.40 | 167908.96 |
| | 0.45 | 183371.28 |
| | 0.50 | 199062.79 |
| | 0.60 | 230936.39 |
| | 0.70 | 263238.31 |
| | 0.80 | 295795.68 |
| | 0.90 | 328506.70 |
| | 1.00 | 361311.27 |

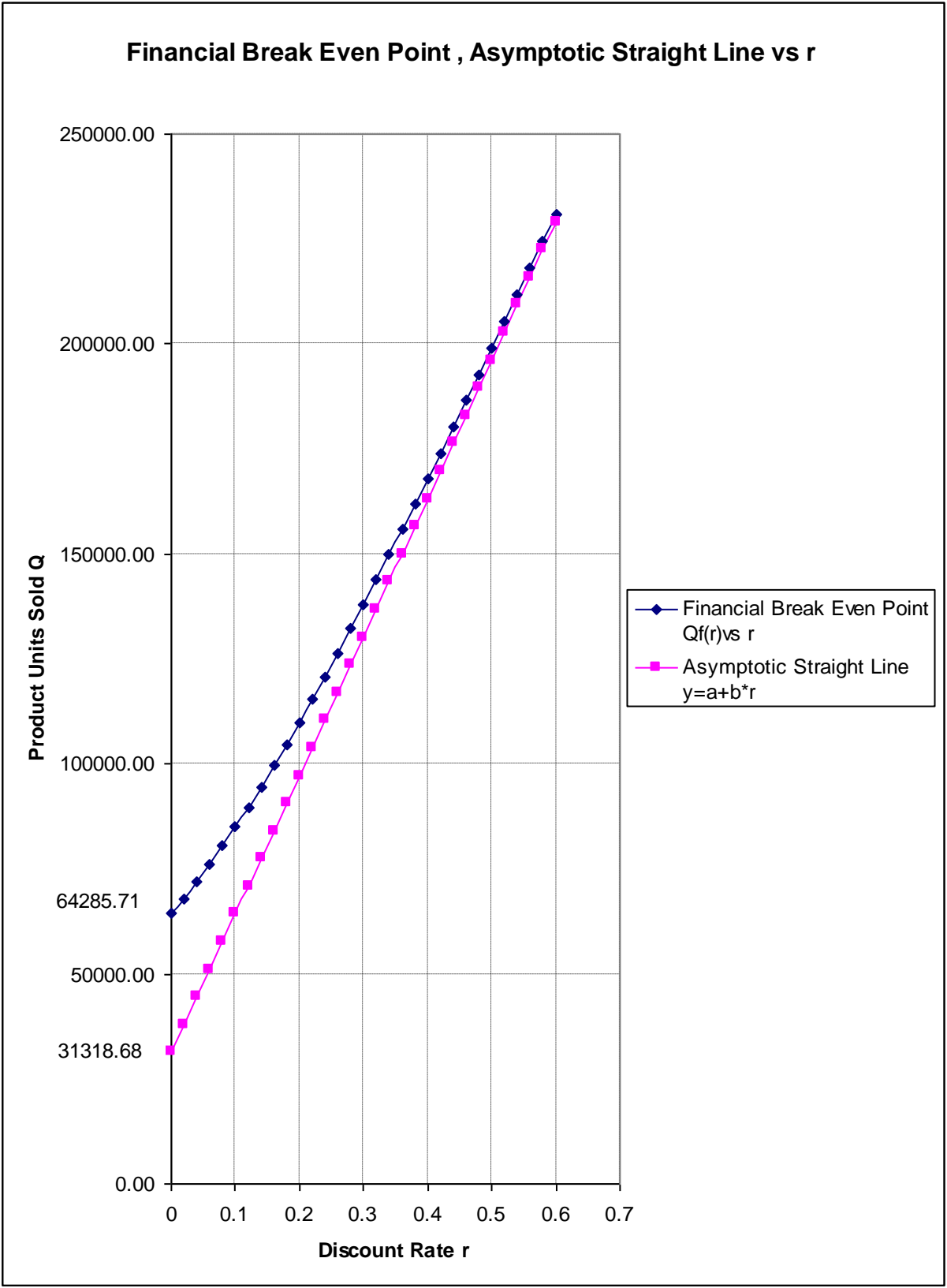

*Figure 2  Graph of* $Q_f(r)$ *vs.* $r$, *and* $y = a + b\,r$ *vs.* $r$

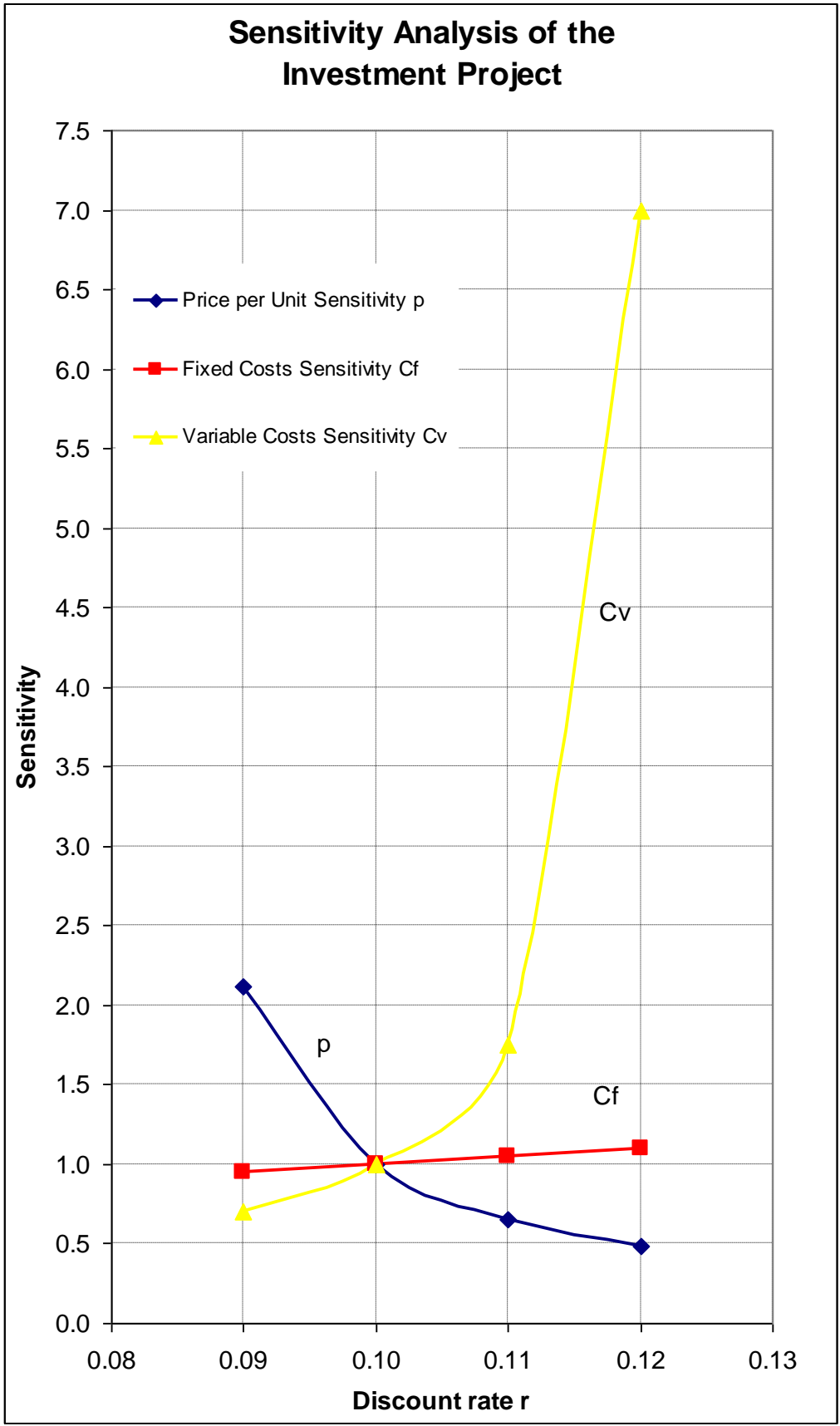

*Figure 3 Sensitivity analysis of* $Q_f$ *with respect to* $C_v, C_f$ *and p*

## 4. Investment Project with the Independent Variable Q and an Annual Growth Rate g

We consider that the investment project defined in Section 2 has a constant annual growth rate $g > 0$ in each year. In this case, we obtain the results given in the next theorem.

*Theorem 6*

The investment project, defined previously, with a annual growth rate $g > 0$ has the following properties:
(i) The $NPV$, as a function of the three variables $Q, r$ and $g$, is given by:

$$NPV(Q,r,g) = h(r) + m(r,g) \, Q, \tag{53}$$

where

$$m(r,g) = (p - C_v)(1 - t_e) \Phi(r,g) > 0, \tag{54}$$

$$\Phi(r,g) = \frac{1 - \left(\frac{1+g}{1+r}\right)^n}{r - g} = \frac{1}{1+g} f\left(\frac{r-g}{1+g}\right), \tag{55}$$

(ii) The financial break-even point $Q_f$ is given, as a function of the discount rate $r$ and the growing rate $g$, by the following expression:

$$Q_f(r,g) = \frac{I - \left((1-t_e)C_f - t_e A\right) f(r)}{\left((1-t_e)C_f - t_e A\right) f\left(\frac{r-g}{1+g}\right)} \frac{(1+g)}{(p - C_v)(1 - t_e)} \tag{56}$$

$$= [b + a f(r)] \frac{1+g}{f\left(\frac{r-g}{1+g}\right)} = (1+g) F\left(\frac{r-g}{1+g}\right)\left(b + \frac{a}{F(r)}\right)$$

where the real functions $f = f(r)$ and $F = F(r)$, and the coefficients $a$ and $b$ are given by the expressions (4), (13), (8) and (9) respectively.

(iii) The financial break-even point $Q_f = Q_f(r,g)$, given by (55), is a strictly increasing function of the discount rate $r$ and it has the following properties:

$$Q_f(0^+, g) = \frac{ng}{(1+g)^n - 1} Q_c < Q_c, \quad Q_f(+\infty, g) = +\infty, \quad \forall g > 0, \tag{57}$$

$$\frac{\partial Q_f}{\partial r}(r,g) > 0, \quad \forall r > 0, \quad \forall g > 0. \tag{58}$$

Moreover, for each $g > 0$, the curve $y = Q_f(r,g)$ vs $r$ has at $r = +\infty$ an asymptotic straight line given by the equation:

$$y = a + b \, r \tag{59}$$

which has a slope $b > 0$ and y-intercept $a$, defined by (9) and (8) respectively. The asymptotic straight line is independent of the growth rate $g > 0$ and coincides with the straight line for the particular case $g = 0$.

*Proof.*

We follow the method developed in Section 2. ∎

## Conclusions

For a simple investment project an explicit expression of the corresponding net present value ($NPV$) as a function of the independent variable $Q$ in order to obtain a closed expression of the financial break-even point $Q_f$ (i.e. the value of $Q$ for which $NPV$ is zero) as a function of the parameters $I$, $n$, $C_v$, $C_f$, $t_e$, $r$, $p$ is derived. The behavior of $Q_f$ as a function of the discount rate $r$ is studied and it is proved that: (i) For $r$ negligible $Q_f$ equals the accounting break-even point $Q_c$ (i.e. the earnings before taxes (EBT) is null) ; (ii) When $r$ is large the graph of the function $Q_f = Q_f(r)$ has an asymptotic straight line with positive slope; (iii) From a sensitivity analysis we conclude that, while the influence of $p$ and $C_v$ on $Q_f$ is strong, the influence of $C_f$ on $Q_f$ is weak; (iv) Moreover, if we assume that the output grows at the annual rate $g$ the previous results still hold, and, of course, the graph of the function $Q_f = Q_f(r, g)$ vs $r$ has, for all $g > 0$, the same asymptotic straight line when $r \to \infty$ as in the particular case with $g=0$.

## Acknowledgements

The present work has been presented at the World Finance Conference, Buenos Aires, 22-24 July 2015, and it has been sponsored by the Projects PIP N° 0534 from CONICET–Universidad Austral.